\documentclass[fleqn,usenatbib]{mnras}

\usepackage{newtxtext,newtxmath}

\usepackage[T1]{fontenc}
\usepackage{ae,aecompl}

\usepackage{graphicx}	
\usepackage{amsmath}	
\usepackage{amssymb}	

\title[A FRB in the direction of Virgo cluster]{A fast radio burst in the direction of the Virgo cluster}

\author[Agarwal et al. (2019)]{
Devansh Agarwal$^{1,2}$\thanks{E-mail: da0017@mix.wvu.edu (DA)},
Duncan R. Lorimer$^{1,2}$,
Anastasia Fialkov$^{3,4,5}$,\newauthor
Keith W. Bannister$^{6}$,
Ryan M. Shannon$^{7}$,
Wael Farah$^{7}$,
Shivani Bhandari$^{6}$,\newauthor
Jean-Pierre Macquart$^{8}$,
Chris Flynn$^{7}$,
Giuliano Pignata$^{10,11}$,
Nicolas Tejos$^{12}$,\newauthor
Benjamin Gregg$^{8}$,
Stefan Os{\l}owski$^{7}$,
Kaustubh Rajwade$^{9}$,
Mitchell B. Mickaliger$^{9}$,\newauthor
Benjamin W. Stappers$^{9}$,
Di Li$^{13,14}$,
Weiwei Zhu$^{13}$,
Lei Qian$^{13}$,
Youling Yue$^{13}$,\newauthor
Pei Wang$^{13}$ and
Abraham Loeb$^{15}$
\\
$^{1}$West Virginia University, Department of Physics and Astronomy, P. O. Box 6315, Morgantown, WV, USA\\
$^{2}$Center for Gravitational Waves and Cosmology, West Virginia University, Chestnut Ridge Research Building, Morgantown, WV, USA\\
$^{3}$Kavli Institute for Cosmology, University of Cambridge, Madingley Road, Cambridge CB3 0HA, UK\\
$^{4}$Institute of Astronomy, University of Cambridge, Madingley Road, Cambridge CB3 0HA, UK\\
$^{5}$Department of Physics and Astronomy, University of Sussex, Falmer,  Brighton BN1 9QH, UK\\
$^{6}$CSIRO Astronomy and Space Science, Australia Telescope National Facility, Box 76 Epping, NSW 1710, Australia\\
$^{7}$Centre for Astrophysics and Supercomputing, Swinburne  University of Technology, Hawthorn, VIC 3122, Australia\\
$^{8}$Department of Astronomy, University of Massachusetts Amherst, Amherst, MA, USA\\
$^{9}$Jodrell Bank Centre for Astrophysics, University of Manchester, Oxford Road, Manchester M13 9P\\
$^{10}$Departamento de Ciencias Fisicas, Universidad Andres Bello, Avda. Republica 252, Santiago, Chile\\
$^{11}$Millennium Institute of Astrophysics (MAS), Nuncio Monse\~{n}or Sotero Sanz 100, Providencia, Santiago, Chile\\
$^{12}$Instituto de F\'isica, Pontificia Universidad Cat\'olica de Valpara\'iso, Casilla 4059, Valpara\'iso, Chile\\
$^{13}$CAS Key Laboratory of FAST, National Astronomical Observatories, Chinese Academy of Sicences,  Beijing 10010, China\\
$^{14}$NAOC-UKZN Computational Astrophysics Centre, University of KwaZulu-Natal, Durban, 4000, South Africa \\
$^{15}$Astronomy Department, Harvard University, 60 Garden St., Cambridge, MA 02138, USA\\
}

\date{Accepted 2019 September 11. Received 2019 September 9; in original form 2019 June 26
}

\pubyear{2019}

\begin{document}
\label{firstpage}
\pagerange{\pageref{firstpage}--\pageref{lastpage}}
\maketitle

\begin{abstract}
The rate of fast radio bursts (FRBs) in the direction of nearby galaxy clusters is expected to be higher than the mean cosmological rate if intrinsically faint FRBs are numerous. In this paper, we describe a targeted search for faint FRBs near the core of the Virgo cluster using the Australian Square Kilometer Array Pathfinder telescope. During 300~hr of observations, we discovered one burst, FRB~180417, with dispersion measure DM~$=474.8$~cm$^{-3}$~pc. The FRB was promptly followed up by several radio telescopes for 27~h, but no repeat bursts were detected. An optical follow-up of FRB~180417 using the PROMPT5 telescope revealed no new sources down to an $R$-band magnitude of 20.1. We argue that FRB~180417 is likely behind the Virgo cluster as the Galactic and intracluster DM contribution are small compared to the DM of the FRB, and there are no galaxies in the line of sight. The non-detection of FRBs from Virgo  constrains the faint-end slope, $\alpha<1.52$ (at 68\% confidence limit), and the minimum luminosity, $L_{\rm min}\gtrsim 2\times 10^{40}$~erg~s$^{-1}$ (at 68\% confidence limit), of the FRB luminosity function assuming cosmic FRB rate of $10^4$ FRBs sky$^{-1}$ day$^{-1}$ with flux above 1 Jy located out to redshift of 1. Further FRB surveys of galaxy clusters with high-sensitivity instruments will tighten the constraints on the faint end of the luminosity function and, thus, are strongly encouraged.
\end{abstract}




\section{Introduction} 

Fast Radio Bursts (FRBs) are extremely bright, highly dispersed pulses of as yet unknown origin. Following the serendipitous discovery of the prototypical ``Lorimer burst'' in archival pulsar search data \citep{Lorimer2007}, FRBs were subsequently confirmed as a cosmological population in dedicated surveys \citep{thornton2013} and now over $\sim$90 sources are known \citep{frbcatalog}\footnote{\url{http://www.frbcat.org}}. Notable discoveries also include the 11 repeating sources: FRB~121102 \citep{2016Natur.531..202S}, nine sources reported by \citet{abb+19b,Andersen2019}, and FRB1701019 \citep{Kumar19}. 
While many theories have been proposed \citep[for a recent compilation, see][]{2018arXiv181005836P}, the origin of both repeating and non-repeating FRBs is still a mystery.

While on-going blind large-area surveys are providing valuable insights into the population \citep{Shannon2018,James2018,James2019}, targeted searches can also prove fruitful. Recently, in one such attempt to optimise searches \citet{Fialkov2018} predict a possible enhancement in the FRB rate in the direction of nearby galaxy clusters if the intrinsically faint FRB population is abundant. Their study was motivated by the availability of small ($\sim 20$~m class) radio telescopes which often have large amounts of observing time available with a modest ($\sim 1$~deg$^2$) field of view, but it can also be investigated by facilities with broader sky coverage. Motivated by these predictions, and the great success of the Australian Square Kilometre Array Pathfinder \citep[ASKAP;][]{Schinckel2012} in finding FRBs \citep{Bannister2017,Shannon2018}, we have conducted a 300~hr survey with ASKAP to look for such an excess in the direction of the Virgo galaxy cluster.

The search was successful in that we found one new FRB~180417 $\sim 3^\circ$ away from the cluster center. In this paper, we describe the survey observations and the properties of this new FRB in \S 2. We also summarise the follow-up observations for repeat bursts in \S 3. In \S 4, we comment on its possible location behind the Virgo cluster. We employ the non-detection of the FRB from the Virgo cluster to derive constraints on the slope and the minimum luminosity cut-off of the FRB luminosity function at the faint-end in \S 5.
\begin{figure*}
\begin{center}
\includegraphics[width=0.9\textwidth]{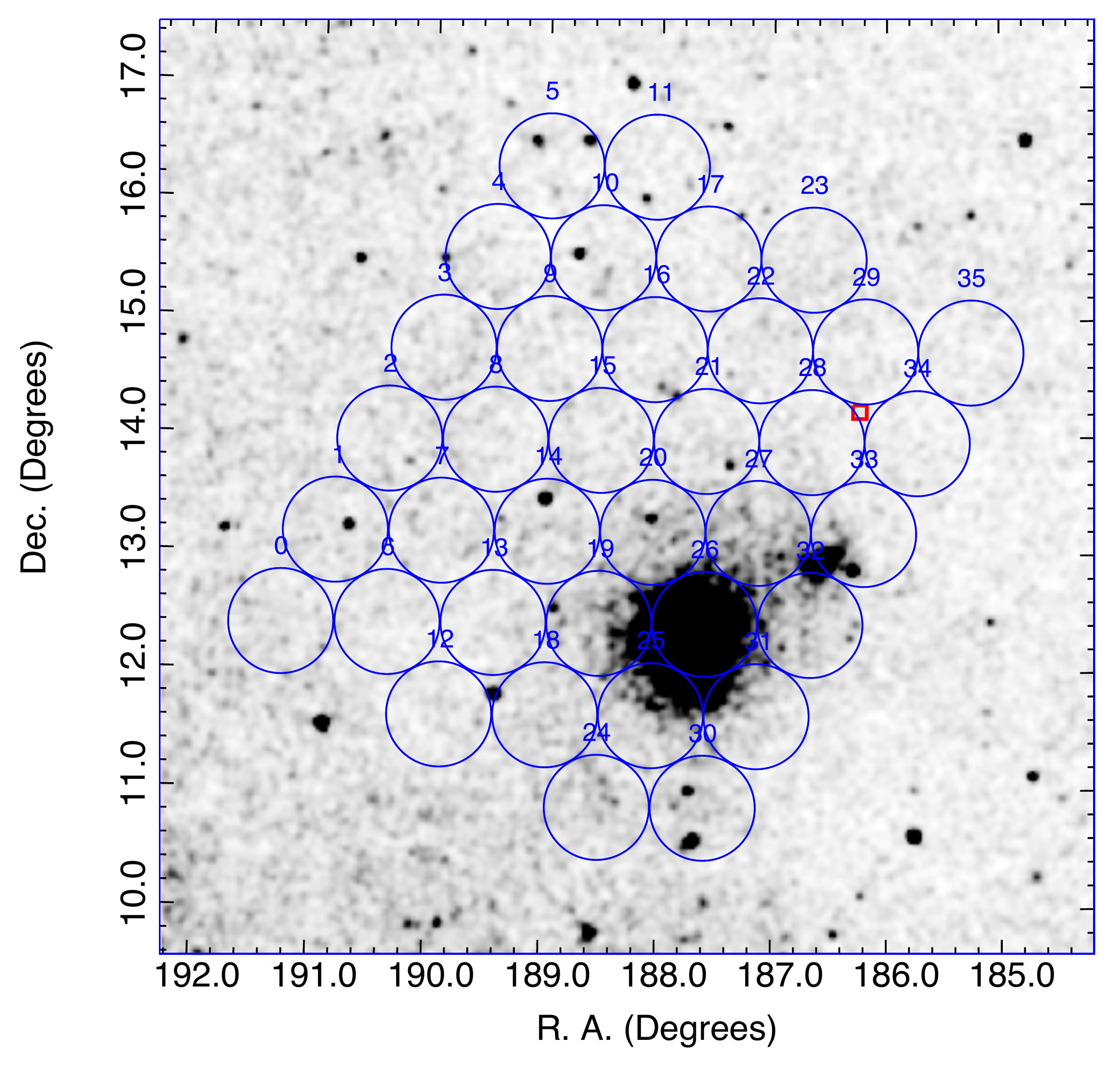}
\caption{The ASKAP footprint overlayed on the ROSAT All-Sky grey scale image of the Virgo cluster of galaxies. The red box denotes the location of FRB~180417. The dark region near Beam 26 is dominated by M87, a giant elliptical galaxy the center of the Virgo cluster.
\label{fig:virgo_rosat}}
\end{center}
\end{figure*}
\section{Observations}
The observations were carried out using the commissioning array under the Commensal Real-Time {ASKAP} Fast-Transients {(CRAFT)} survey \citep{Macquart2010}. Depending on availability we used 6--8 ASKAP antennas in the incoherent summed mode. The observations were carried out from March 9, 2018, to May 9, 2018, with approximately seven hours per day. The field center is right ascension (RA) $12$h$33$m and declination (Dec) $+13$d$34$m in the J2000 epoch. These coordinates were reported by \citet{Fialkov2018} for the maximum FRB rates from Virgo. Fig.~\ref{fig:virgo_rosat} shows the ASKAP footprint overlayed on a ROSAT image of the cluster \citep{Truemper1982}. The data capturing pipeline is detailed in \citet{Bannister2017}. Total intensity streams from 36 beams of each antenna were recorded on the disk and summed offline. The data were then searched for FRBs using the identical pipeline as described in \citet{Bannister2017}. We use the graphics processing unit accelerated real-time search pipeline FREDDA (Bannister et al. (in prep)) and search for 12 different pulse widths in the range $1.26$--$15.12$~ms over a dispersion measure (DM) interval of $20$--$4096$~cm$^{-3}$~pc. Candidates were clustered together using the friends of friends of algorithm \citep{Huchra1982} and archived along with their maximum signal to noise ratio (S/N). Clustered candidates with S/N~$>10$ were selected for subsequent visual inspection.

\section{Results}
One FRB was detected as a result of these observations and data processing, FRB~180417. We detail the parameters of this source and the follow-up observations we carried out in the subsections below.

\subsection{FRB~180417}
\begin{figure}[h!]
\begin{center}
\includegraphics[width=0.65\columnwidth]{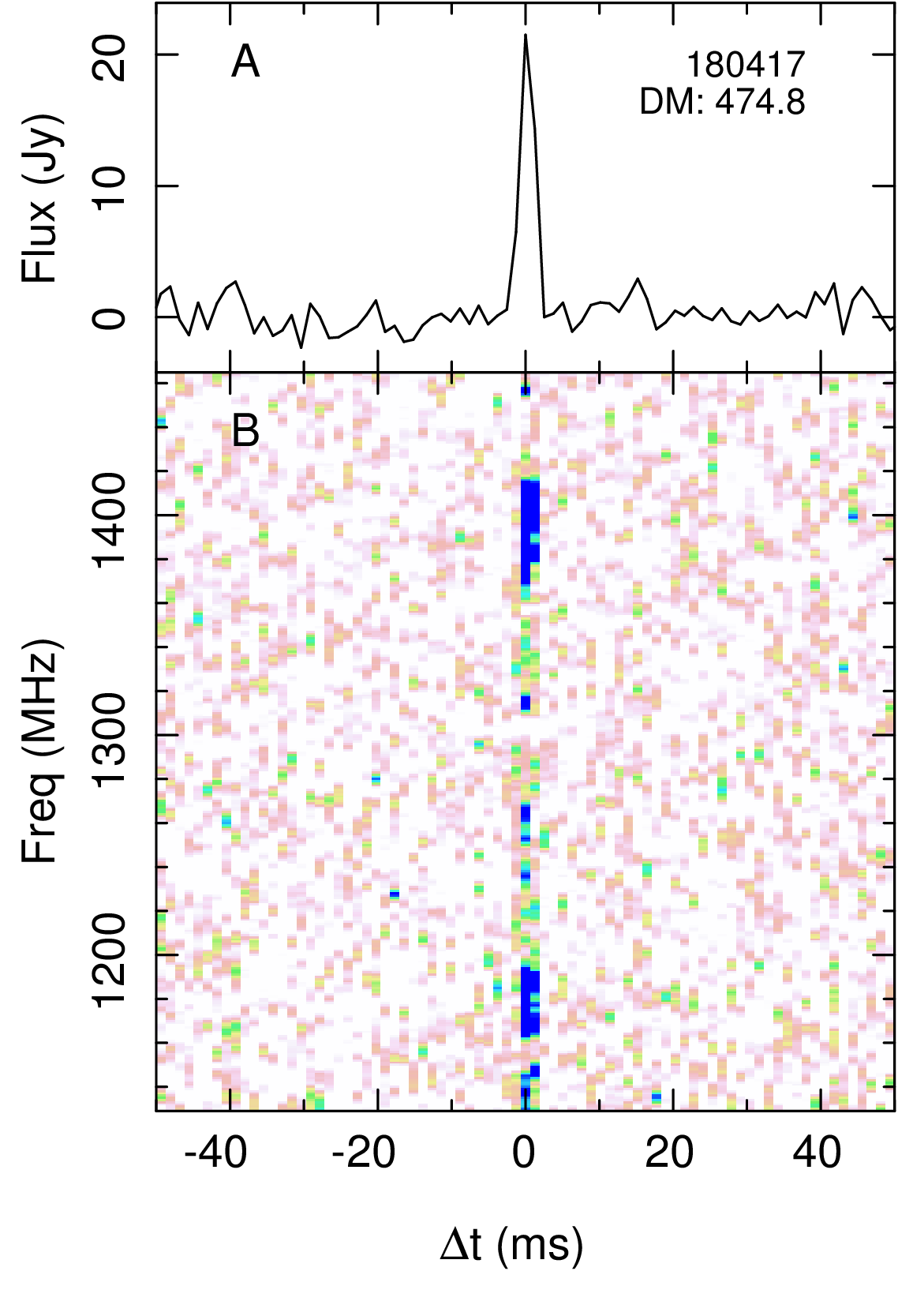}
\caption{The dedispersed profile and dynamic spectrum for FRB~180417. The top panel shows the co-added profile from all three beams. The bottom panel shows the dynamic spectrum of the FRB. The frequency structure of the FRB is clumpy which is similar to previously reported FRBs from ASKAP \citep{Macquart2019}.}
\label{fig:dyn_spec}
\end{center}
\end{figure}

FRB~180417 was strongly detected in three beams with S/N $>14$, and in a further two beams with (S/N $> 5$), as shown in Table \ref{tab:0}.

Fig.~\ref{fig:dyn_spec} shows the frequency versus time plot with  S/N $=24.2$ from the co-addition of these beams. The pulse was detectable at S/N~$\sim 5$ in individual antennas with similar frequency structure.

The estimated Galactic DM contribution in the direction of the FRB using NE2001 \citep{ne2001I,ne2001II} and YMW16 \citep{Yao2017} is 26.15~pc~cm$^{-3}$ and 20.39~pc~cm$^{-3}$ respectively. We estimate the Galactic halo DM contribution to be $\approx$~30~pc~cm$^{-3}$ (see \citet{Dolag2014} and section \S~\ref{subsec:4.1} for more discussion). Properties of the FRB are summarised in Table~\ref{tab:1}.
 \begin{table}
 \centering
 \caption{Detection S/N of FRB~180417}
 \label{tab:0}
\begin{tabular}{cccr}
\hline
Beam  & RA (J2000) & DEC (J2000) & S/N \\
\hline
 21 &  12:30:20  & 13:58:07 & 0.6 \\
22 &  12:28:28  & 14:44:48 & 5.4  \\
27 &  12:28 30  & 13:11:15 & 3.0  \\
28 &  12:26:38  & 13:57:52 & 15.0 \\
29 &  12:24:45  & 14:44:25 & 16.8 \\
33 &  12:24:48  & 13:10:44 & 5.9 \\
34 &  12:22:55  & 13:57:24 & 14.0 \\
\hline
\end{tabular}
\end{table}
\begin{table}
\centering
\caption{Observed properties of FRB~180417}
\label{tab:1}
\begin{tabular}{cc}
\hline
Parameter & Value \\
\hline
UTC & 2018-04-17 13:18:31 (at 1297~MHz)\\
MJD & 58225.55452546\\
S/N & 24.2 \\
DM  & 474.8~pc~cm$^{-3}$ \\
RA (J2000) & 12h~24m~56(28)s  \\
Dec (J2000) & +14d~13(7)m\\
Boxcar Width & 2.52~ms\\
Fluence & 55(3)~Jy~ms\\
\hline
\end{tabular}
\end{table}
The multiple-beam detections of FRB~180417 allow us to constrain the burst location and fluence. 
To do so, we use the method described in detail in \S~4.1 of \citet{Bannister2017} which we summarise here. Using a model for the responses for adjacent beams, we use the beam positions on the sky and burst S/N to infer the burst position and attenuation. 
The position and attenuation are inferred using Bayesian methodology, after accounting for uncertainties in beam gain, shape and position. 
The method has been found to be robust in bursts with the position derived for FRB~180924 using this method consistent with the interferometric position \cite[][]{Bannister19}, and with the detection of repeat pulses of ASKAP FRB~171019 with the Green Bank telescope \cite[][]{Kumar19}.  
Using the positions and S/N for the beams around the FRB~180417 detection (see Table \ref{tab:0}), 
we are able to constrain the location to an error box of size 7$^\prime \times$ 7$^\prime$ and the fluence as 55 $\pm$ 3 ~Jy~ms.

\begin{figure}
\begin{center}
\includegraphics[width=\columnwidth]{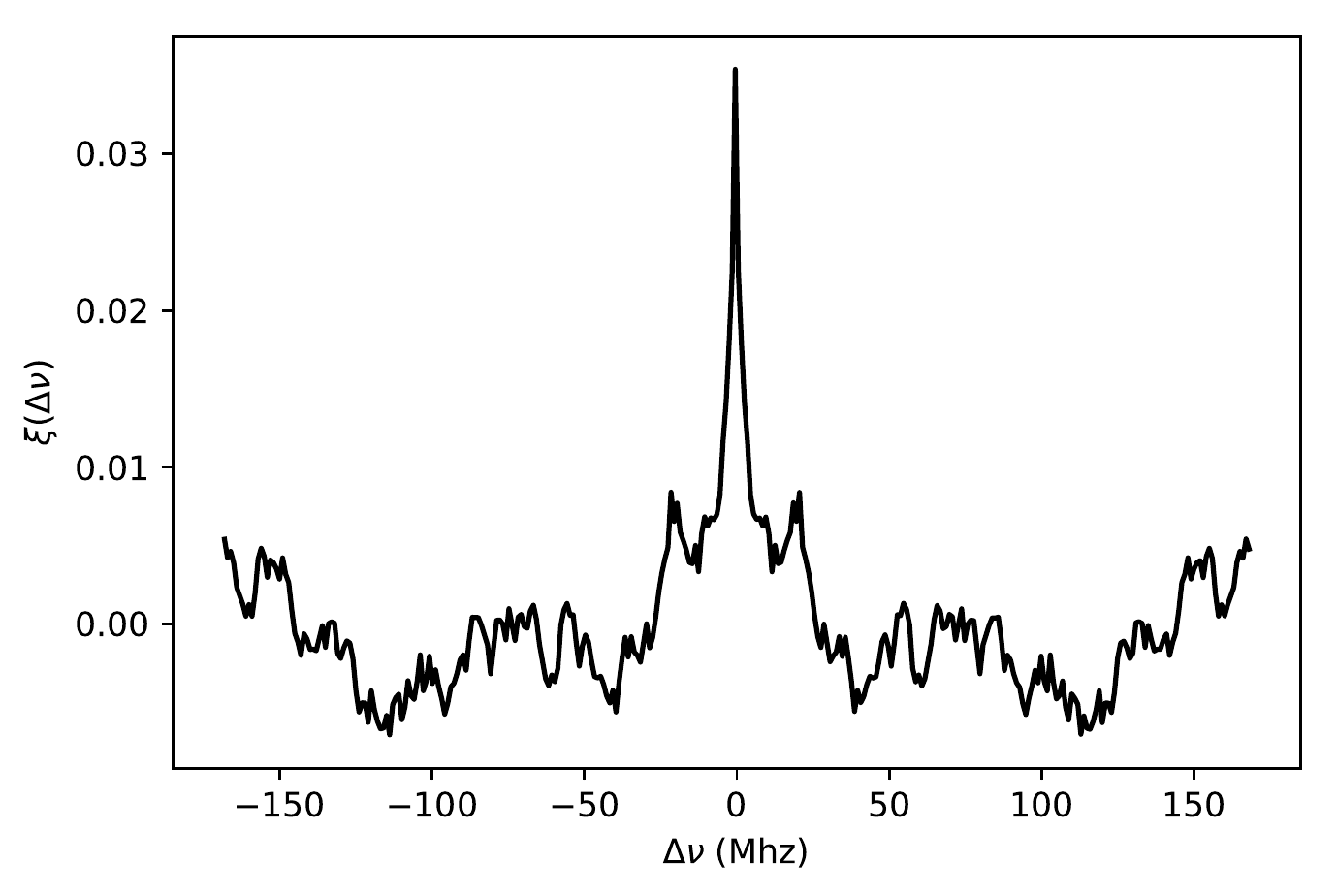}
\caption{The auto-correlation function of the spectrum of FRB180417.} 
\label{Fig:ACF}
\end{center}
\end{figure}
We characterise the spectral variations by computing the mean normalised auto-correlation function of the spectrum ($f_\nu$) as
\begin{equation}
    \xi (\Delta \nu) = \frac{\big \langle [f_\nu(\nu^\prime + \Delta \nu) -\Bar{f}_\nu] [f_\nu(\nu^\prime) -\Bar{f}_\nu] \big \rangle}{\Bar{f}_\nu^2}.
\end{equation}
Here $\Bar{f_\nu}$ is the mean spectrum amplitude. Figure \ref{Fig:ACF} shows the auto-correlation function of the FRB spectra. We fit the above with $\xi (\Delta \nu) = m /(f_{\rm dc}^2 + \Delta \nu^2)$ and obtain a decorrelation bandwidth, $f_{\rm dc} = 4.3 \pm 0.4$~MHz and the modulation index, $m = 0.47 \pm 0.07$ \citep{Cordes1986}. This is consistent with expectation for the ISM at this location on the sky based on the NE2001 model. The NE2001 model estimates $f_{\rm dc, NE2001} = 6.3$~MHz at 1.4~GHz, implying the ISM is responsible for the spectral variations.

\subsection{Radio follow-up observations}

We have undertaken an extensive follow-up campaign to search for repetition from FRB~180417. Owing to the nature of our survey, we have repeatedly covered the region of FRB~180417. The FRB was discovered when 53\% of our 300~hr survey was completed. We have spent a total of 27.1 hours searching at the location of the burst with other telescopes as detailed below.

Starting soon after the detection, we began following up using various other telescopes. The most rapid follow-up occurred with the Parkes and Lovell radio telescopes which were able to perform a search for repeated bursts within 24 hours of the original detection, with the The Five-hundred-meter Aperture Spherical radio Telescope {(FAST)} and a 20-m dish at the Green Bank Observatory joining soon after. The advantage of the follow-up using larger telescopes is the increased sensitivity which is beneficial as we expect there would be weaker bursts, in line with the observed properties of FRB~121102. Under our follow-up, FAST was the most sensitive telescope with 0.03~Jy~ms fluence limit \citep{NAN2011, Li2018}. The data were searched for DM range or 400--550~pc~cm$^{-3}$ with 1000 trials using {\sc heimdall}\footnote{\url{https://sourceforge.net/projects/heimdall-astro}}. Candidates with S/N~$>6$ were inspected visually. Table \ref{tab:2} describes the follow-up details. We did not detect any repeat bursts, and we defer detailed limits and modelling to a separate publication.

\begin{table}
	\centering
	\caption{Details of the radio follow-up of FRB~180417. Here F$_\mathrm{min}$ is the minimum fluence detectable by the telescope.}
    \label{tab:2}
    \begin{tabular}{lcc}
        \hline
        Telescope & Observation Length (hr) & F$_\mathrm{min}$ (Jy~ms)\\
        \hline
        GB~20m &  16.0 & 4.8\\
        FAST & 0.5  & 0.03\\
        Parkes & 6.6 & 2.0\\
        Lovell & 4.0 & 0.5\\
        \hline
    \end{tabular}
\end{table}
\subsection{Optical follow-up}

\begin{figure}
\includegraphics[width=\columnwidth,trim={1.25cm 0.25cm 2cm 1.4cm},clip]{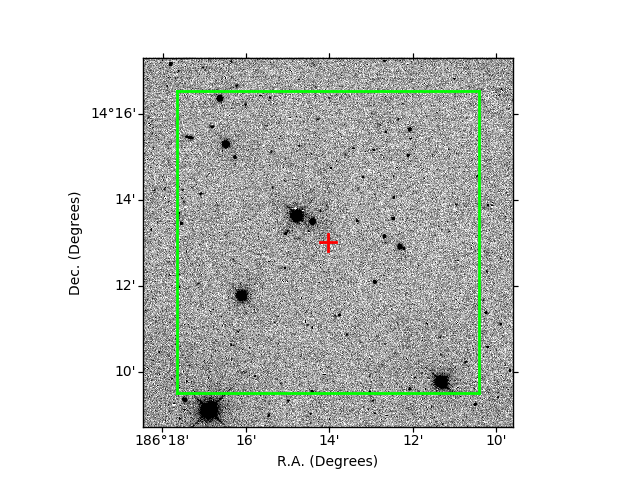}
\caption{PROMPT5 image acquired on 2018 May 11.96 UT. The red cross indicates the FRB~180417 position, while the green square shows the corresponding error box.  
\label{fig:opt_frb}}
\end{figure}

Optical imaging at the location of the FRB~180417 (red cross in Fig.~\ref{fig:opt_frb}) was carried out on 2018 May 11.96 UT with the 40~cm PROMPT5 telescope located at CTIO. PROMPT5 has a field of view of $11^{\prime}\times 11^{\prime}$ fully covering the position uncertainty derived by the ASKAP observations (green box in Fig.~\ref{fig:opt_frb}). A series of thirty 40~s $R$-band images were acquired for a total integration time of 20~min. Each frame was correct for bias, dark and flat using standard routines in IRAF. A final image was obtained taking a median value for each pixel. The photometry was calibrated using the magnitude of stars present in the PROMPT field of view, reported in the Pan-STARRS photometric catalog \citep{Magnier2016} transformed to the Johnson Kron-Cousins photometric system using the transformation reported in \citet{Smith2002}.

To search for an optical counterpart FRB~180417, we searched optical archives looking for images obtained before the FRB occurrence. In the Canada France Hawaii Telescope (CFHT) archive we found an $r$ band MegaCam image with a total integration time of 1374~s acquired on 2013 May 14th, which fully covered the PROMPT5 image. We aligned, re-scaled and convolved the MegaCam image with SWarp \citep{Bertin2002} and  HOTPANTS \citep{Becker2015} in order to match the orientation, flux and PSF of the PROMPT5 frame.

In the template subtracted image, we searched for transients using algorithms developed for the CHASE survey \citep{Pignata2009}. We did not detect any source with S/N~$>3$. Using artificial stars placed around the FRB~180417 position, we set an upper limit of $R = 20.1$ on the optical counterpart detection. The small blank regions in the MegaCam mosaic are covered by one of the sub-frames of a $R$ band VMOS image acquired on 2009 February 26th, we found in the ESO archive, which has an integration time of 180~s. We use the latter image as a template in the same way we did for the MegaCam frame, however,  no sources with S/N~$>3$ were detected.

\section{Implications of FRB~180417}

\subsection{Is FRB~180417 in the Virgo cluster?}
\label{subsec:4.1}
To estimate the distance of the FRB we perform a simple analysis in which the DM of FRB~180417, {DM}$_\mathrm{FRB}$, is represented as the sum of contributions from the Milky Way (MW), intracluster medium (ICM), intergalactic medium (IGM), and the host, as follows:
\begin{equation}
    \text{DM}_\mathrm{FRB} = \text{DM}_\mathrm{MW} + \text{DM}_\mathrm{ICM} + \text{DM}_\mathrm{IGM} + \text{DM}_\mathrm{host}.
\end{equation}
Using two different Galactic electron density models NE2001 \citep{ne2001I,Yao2017} for this line of sight, and taking a 
Galactic halo contribution of 30~cm$^{-3}$~pc, we find DM$_\mathrm{MW}=60$~cm$^{-3}$~pc. Virgo is at a redshift of $\approx0.004$, and the contribution to the DM due the IGM is expected to be $\approx 5$~cm$^{-3}$~pc \citep[using DM redshift relation from][]{Inoue2004}, and is considered to be negligible. We model DM$_\mathrm{ICM}$ from the Virgo cluster as described below. Lastly, we leave the host galaxy contribution, DM$_\mathrm{host}$ as a free parameter.

\citet{Ade2016} have used X-ray and Planck data to estimate the electron density ($n_e$) out to two viral radii (2.4~Mpc) as a function of the radius. Using this model, the electron density is, 
\begin{equation}
    n_e(b,z_\mathrm{LOS}) = \frac{8.5\times10^{-5}\mathrm{cm}^{-3}}{(b^2 + z_\mathrm{LOS}^2)^{0.6}}.
\end{equation}
Here, $z_\mathrm{LOS}$ is the depth along the line of sight (not to be confused with the redshift) and $b$ is the impact parameter, both in Mpc. $z_\mathrm{LOS}=0, b=0$ corresponds to M87, the center of the cluster. 
FRB~180417 is located $2.3^\circ$ from the center of the cluster which corresponds to $b\sim 0.67$~Mpc corresponding to 0.55 times the virial radius. As a result, the intracluster contribution is
\begin{eqnarray}
    \text{DM}_\mathrm{ICM} &=& 10^6\, \text{cm}^{-3}~\text{pc} \int_{-2.4}^{2.4} n_e(b=0.67,z_{\mathrm{LOS}})\, d z_{\mathrm{LOS}}\\
                           &=& 332 \, \text{cm}^{-3}~\text{pc}. \nonumber
\end{eqnarray}
If FRB~180417 is indeed in the Virgo cluster, then we can place a lower bound on the DM$_\mathrm{host}$ to be 90~cm$^{-3}$~pc.

The location of FRB~180417 is at the outskirts of Virgo, where galaxy crowding is low. According to the Virgo catalog \citep{Binggeli1985,Kim2014}, the closest galaxy, EVCC~0548 is a dwarf spiral (dS0) galaxy, 6.3$^\prime$ away on the sky from the line of sight to FRB~180417 and has half-light radius of 7.5$^{\prime\prime}$. The next nearest galaxy is EVCC~0567 which has dwarf elliptical morphology, is 12$^{\prime}$ away from the FRB location and has half-light radius of 24$^{\prime\prime}$. For both of the galaxies, there are no counterparts in the NVSS catalog. Hence, it is difficult to associate the FRB with a member galaxy of Virgo.

\subsection{Probing the Virgo Intra-cluster Medium}

Assuming that the FRB occurred behind the Virgo cluster, we can probe the intracluster (ICM) medium by placing constraints on scatter broadening of the pulse profile. Turbulence in the ICM would cause the radio pulse to diffract, which, if sufficiently strong would cause the pulse to temporally smear. In the case of this observation, we assume that the pulse is emitted at a distance much further than the Virgo cluster, so that we can assume the signal has a plane parallel geometry at the Virgo cluster. Following Eq. 9 from \cite{ne2001I}, the pulse scattering time at the distance of the Virgo Cluster (16.5~Mpc) is,
\begin{eqnarray}
\tau = 5.8 \, {\rm SM}^{6/5} \, \nu^{-22/5} \,\, {\rm s}. 
\end{eqnarray}
Here SM is the scattering measure in its conventional units of kpc\,m$^{-20/3}$, and $\nu$ is the frequency in GHz.
As the pulse width is only two bins, we assume the scatter broadening to be less than a sample i.e.\,1.26 ms.  Assuming pulse scattering to be less than one sample, i.e. $\tau < 1.26$\,ms, we find ${\rm SM} < 10^{-3.06}\,$kpc\,m$^{-20/3}$, which can be expressed in terms of the root mean square of the electron column along the line of sight at the outer scale of the turbulence, $L_0$ of 
\begin{eqnarray}
\langle \Delta {\rm DM}^2 \rangle^{1/2} = 1.95 \, \left( \frac{L_0}{1\,{\rm pc}} \right)^{5/6}\, \hbox{pc\,cm}^{-3}.
\end{eqnarray}
This limit is not strongly constraining on the scattering properties of the medium.  

To place this in context, one may crudely approximate the intra-cluster medium as a uniform slab of material extending out to twice the virial radius of 1.2\,Mpc.  This implies a limit on the in situ ``level of turbulence'' of $C_N^2 < 3.7 \times 10^{-7}\,$m$^{-20/3}$ (noting that the scattering measure is the integral of the level of turbulence along the ray path: ${\rm SM} = \int C_N^2(z)  dz$).  One might plausibly expect the value of $C_N^2$ to be considerably lower than the limit found here for the typical plasma densities and turbulence parameters within an intra-cluster environment.  To illustrate this point, consider a medium of mean electron density $\bar N_e$ which gives rise to density fluctuations with variance $\langle \Delta n_e^2 \rangle= \alpha^2 \bar N_e^2$, at some outer scale $L_0$, plausibly of order $\sim 1\,$kpc for the ICM.  This would have a characteristic level of turbulence of  
\begin{eqnarray}
C_N^2 \approx 6.7 \times 10^{-9} \alpha \left( \frac{\bar N_e}{10^{-3}\,{\rm cm}^{-3}} \right)^2 \left( \frac{L_0}{1\,{\rm kpc}} \right)^{-2/3} {\rm m}^{-20/3} ,
\end{eqnarray}
where $\alpha$ is likely of order unity \cite[][]{1988AIPC..174..185A} and we have normalised to fiducial values for an intra-cluster environment.  Thus we observe that the present upper limit on the scattering measure, and in turn $C_N^2$, is still a factor $\sim 50$ above that which might be expected in intra-cluster plasma.

\subsection{The FRB luminosity function}
\begin{figure*}
\begin{center}
\includegraphics[width=\columnwidth]{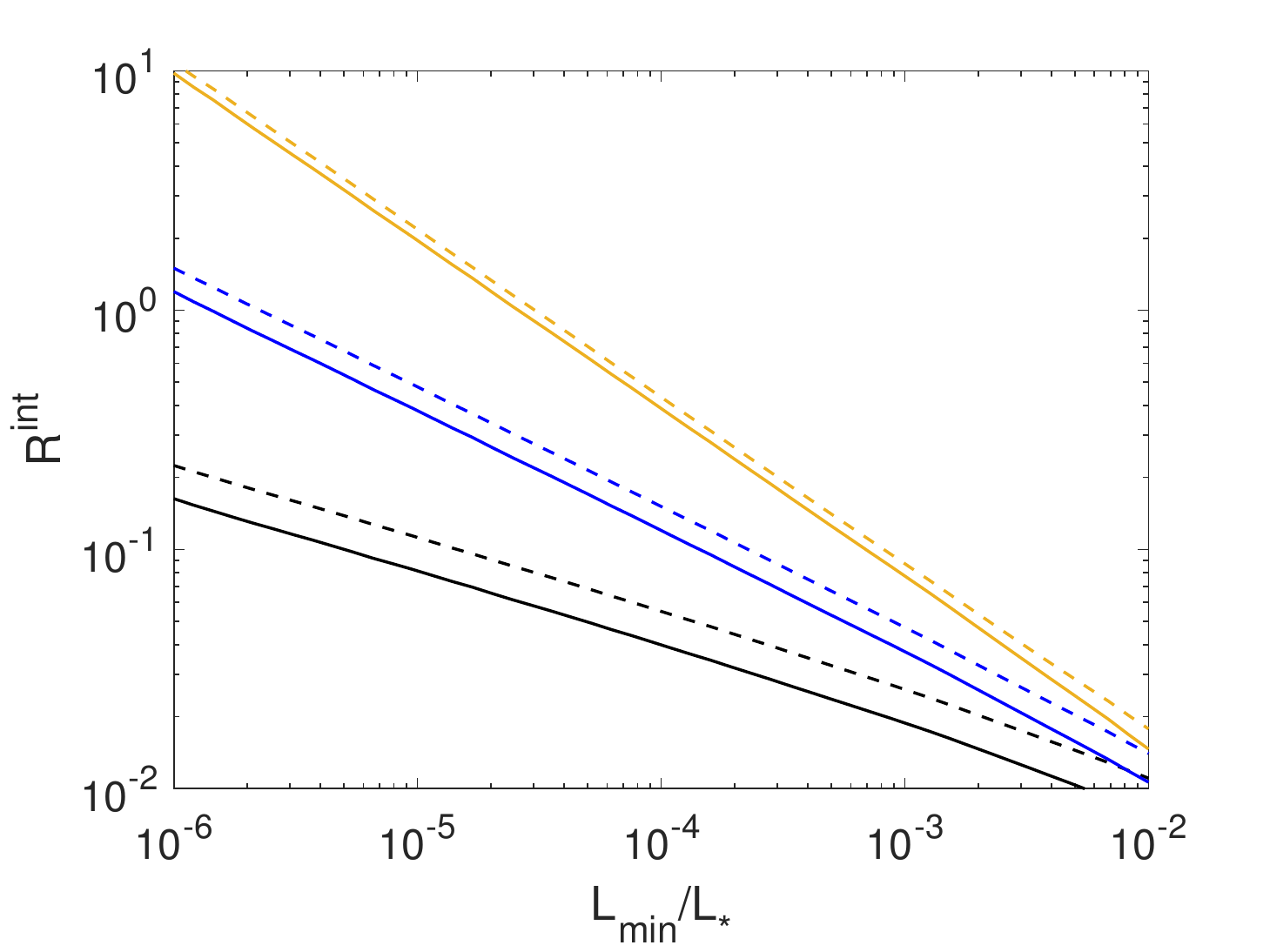}
\includegraphics[width=\columnwidth]{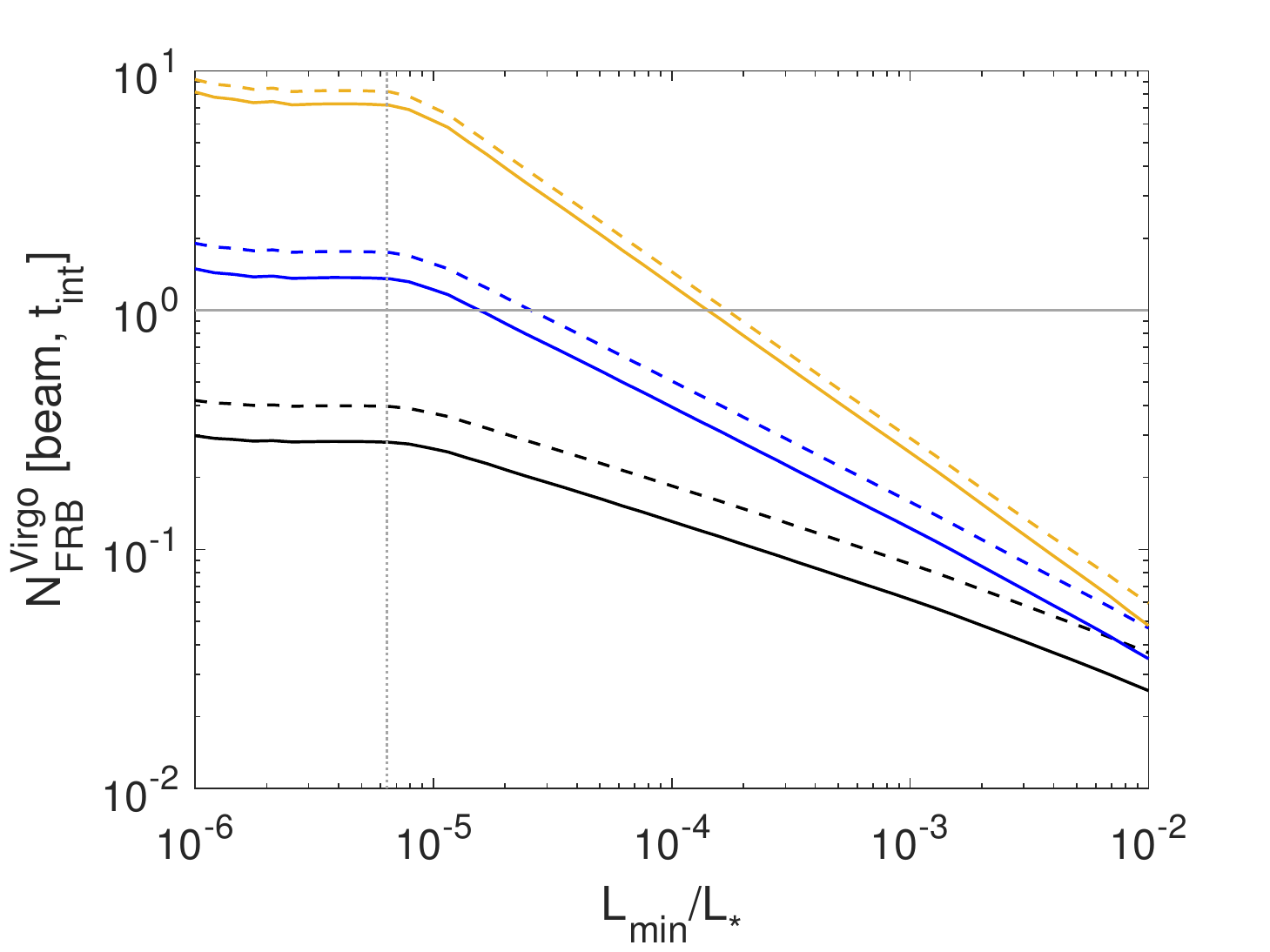}
\caption{Left: normalisation factors, $R_{\rm SFR}^{\rm int}$ (solid) and $R_{M*}^{\rm int}$ (dashed) for Schechter luminosity function with the faint-end slope $\alpha = 1.3$ (black), $\alpha = 1.5$ (blue), $\alpha = 1.7$ (orange) calculated assuming a total of $10^4$ FRBs per sky per day above a detection threshold of 1 Jy and out to a redshift of 1.  Right: expected total number counts from Virgo within ASKAP field of view  over 300 hours integration time calculated using the normalisation coefficients shown on the left panel. Same colour-code is used.
The horizontal line shows $\langle N \rangle =1$ for reference. The vertical dotted line corresponds to the luminosity of the faintest Virgo FRB above the sensitivity limit of ASKAP.} 
\label{Fig:NFRBs}
\end{center}
\end{figure*}
Due to the small number statistics of FRBs, their luminosity function is poorly constrained. Recently, \citet{Luo2018} used 33 FRBs from the online FRB catalog to constrain parameters of the FRB luminosity function assuming the Schechter form so that the differential number of FRBs per unit luminosity interval is
\begin{equation}
\frac{dN_{\rm FRB}}{dL_\nu} \propto \left(\frac{L_\nu}{L_{\nu*}}\right)^{-\alpha}\exp\left[-\frac{L_\nu}{L_{\nu*}}\right], ~ L>L_{\rm min},
\end{equation}
where $\alpha$ is the faint-end slope and $\nu L_{\nu*}$ is the characteristic luminosity of FRBs.  The luminosity function is normalised to unity between the minimum intrinsic luminosity  $L_{\rm min}$ and the maximal brightness (which we assume to be $10 L_{\nu*}$) and plays the role of the probability density of FRB luminosities.  \citet{Luo2018}  found the slope ranging between  $1.2$ to $1.8$ with the best-fit values of $\alpha \sim 1.5$  and $L_* \sim 2\times 10^{44}$ erg~s$^{-1}$. From the sample, it was impossible to measure $L_{\rm min}$ due to the limited number of sources. In addition, random FRB searches typically probe mean cosmological population and pick up intrinsically brighter FRBs located at intermediate cosmological distances. For example, the 20 new FRBs recently reported by \citet{Shannon2018} were detected using ASKAP in the fly's eye mode and are probing the bright-end of the luminosity function.  The survey reported here is unique in that, by surveying the nearby clustered environment of Virgo located only $\sim$16.5~Mpc away, ASKAP can detect faint FRBs down to  $L \sim 1.3\times 10^{39}$ erg~s$^{-1}$ which corresponds to its flux limit  $S_{\rm lim, ASKAP} = 26/\sqrt{7}$~Jy. The factor of $\sqrt{7}$ is due to incoherent sum of data from (on an average) 7 antennas.

The expected FRB number counts from Virgo depend on the shape of the luminosity function, cosmic FRB event rate (used for normalisation),  the nature of the progenitors and the spectral energy distribution of the bursts. In \citet{Fialkov2018} we considered two types of the luminosity function for FRBs: (i) standard candles with fixed luminosity of $\nu L_{\nu*} = 2.8\times 10^{43}$ erg~s$^{-1}$ which corresponds to the mean intrinsic luminosity of the observed FRBs (excluding the recently discovered ASKAP events); and (ii)  the Schechter luminosity function. \citeauthor{Fialkov2018}~showed that if FRBs are standard candles, the contribution of the supercluster is negligible compared to the cosmological contribution within the solid angle of Virgo. However, owing to its proximity, Virgo is expected to dominate the FRB number counts in cases where the faint-end population is numerous (e.g., in the case of a Schechter luminosity function with sufficiently low $L_{\rm min}$ and steep faint-end slope). Assuming that FRB~180417 is outside Virgo, no other FRBs were found in the observed area during the 300 hr survey. Using this information, we can provide new limits on the intrinsically faint population of FRBs constraining $L_{\rm min}$ for the first time.  
 
The procedure is as follows: First, we follow the method outlined in \citet{Fialkov2018} to calculate per-galaxy FRB event rate based on a  cosmological population of FRBs as a function of $\alpha$ and $L_{\rm min}$ and assuming a fixed total rate of  $\dot N_{\rm FRB} = 10^4$~FRBs per sky per day\footnote{Because FRB rates are very uncertain, we also quote the final results for the total of $10^3$ FRBs per sky per day above the detection threshold of 1~Jy out to redshift $z=1$.} above the detection threshold of 1~Jy out to redshift $z=1$ \citep[e.g.,][]{Nicholl2017}.  Next, we apply this rate to Virgo galaxies extracted from an online Virgo catalogue \citep{Kim2014} and calculate the expected number of FRBs within the 300 h survey with ASKAP, $\langle N_{\rm FRB}^{\rm Virgo} \rangle$. Finally, we employ Poisson statistics to assess the probability of  non-detection of FRBs from Virgo and place limits on $\alpha$ and $L_{\rm min}$.

The cosmic event rate is given by
\begin{eqnarray}
\dot N_{\rm FRB} = \int_V dV \int_{M_h} d M_h \frac{d}{dM_h}n(z,M_h)\frac{\dot N_1(z,M_h)}{(1+z)} \nonumber \\
\int_{S>S_{\rm min}} dL\left(\frac{L_\nu}{L_{\nu*}}\right)^{-\alpha}\exp\left[-\frac{L_\nu}{L_{\nu*}}\right]
\end{eqnarray}
where the comoving halo abundance per unit volume ($dn(z, M_h)/dM_h$) is calculated using Sheth-Tormen mass function \citep{Sheth1999}, the $(1 + z)$ factor accounts for cosmological time dilation and $\dot N_1(z,M_h)$ is the FRB rate per halo. $S_{\rm min}$ is the larger of the telescope sensitivity and the observed  flux of the dimmest intrinsic FRB from redshift $z$, given by $L_{\rm min}(1+z)/[4\pi D_L^2(z)]$, and  $D_L(z)$ is the luminosity distance to the FRB. As in \citet{Fialkov2018}, we use two models for the FRB progenitors to relate the per-halo rates to the properties of actual galaxies.
In the first case, we assume that FRBs trace  star formation rate (SFR) and the FRB rate is given by: 
\begin{equation}
\dot N_1(z,M_h) = R^{\rm int}_{\rm SFR}\left( \frac{{\rm SFR}(z,M_h)}{{\rm SFR}_{\rm Virgo}} \right),
\end{equation}
 where $R^{\rm int}_{\rm SFR}$ is the normalisation coefficient fixed to yielda total of $\dot N_{\rm FRB} = 10^4$~FRBs per sky per day above the detection threshold of 1~Jy out to redshift $z=1$, SFR$(z,M_h)$ is the cosmic mean star formation rate in halos of mass $M_h$ at redshift $z$ calculated using the method of \citet{Behroozi:2013} and ${\rm SFR}_{\rm Virgo}=776$~M$_{\odot}$~yr$^{-1}$ is an estimate of the total ${\rm SFR}$ in Virgo \citep[estimated following][]{Fialkov2018}.  In the second scenario, the FRB rate is proportional to the stellar mass $M_*$:
 \begin{equation}
 \dot N_1(z,M_h) = R^{\rm int}_{M*} M_*(z,M_h)/M_{\rm Virgo},     
 \end{equation}
 where $R^{\rm int}_{M*}$ is the normalisation coefficient, $M_{\rm Virgo}$ is the total stellar mass in Virgo $M_{\rm Virgo} \sim 6\times 10^{12}$ M$_\odot$,and $M_*(z,M_h)$ is the total stellar mass in a halo of mass $M_h$ at redshift $z$. $M_*$ and $M_h$ are related via the star formation efficiency which we also adopt from the work by \citet{Behroozi:2013}.  For a fixed cosmic FRB event rate, the normalisation coefficients depend on both $\alpha$ and $L_{\rm min}$ and are shown in Figure \ref{Fig:NFRBs} assuming $\dot N_{\rm FRB} = 10^4$~FRBs per sky per day. 

\begin{figure*}
\begin{center}
\includegraphics[width=\textwidth]{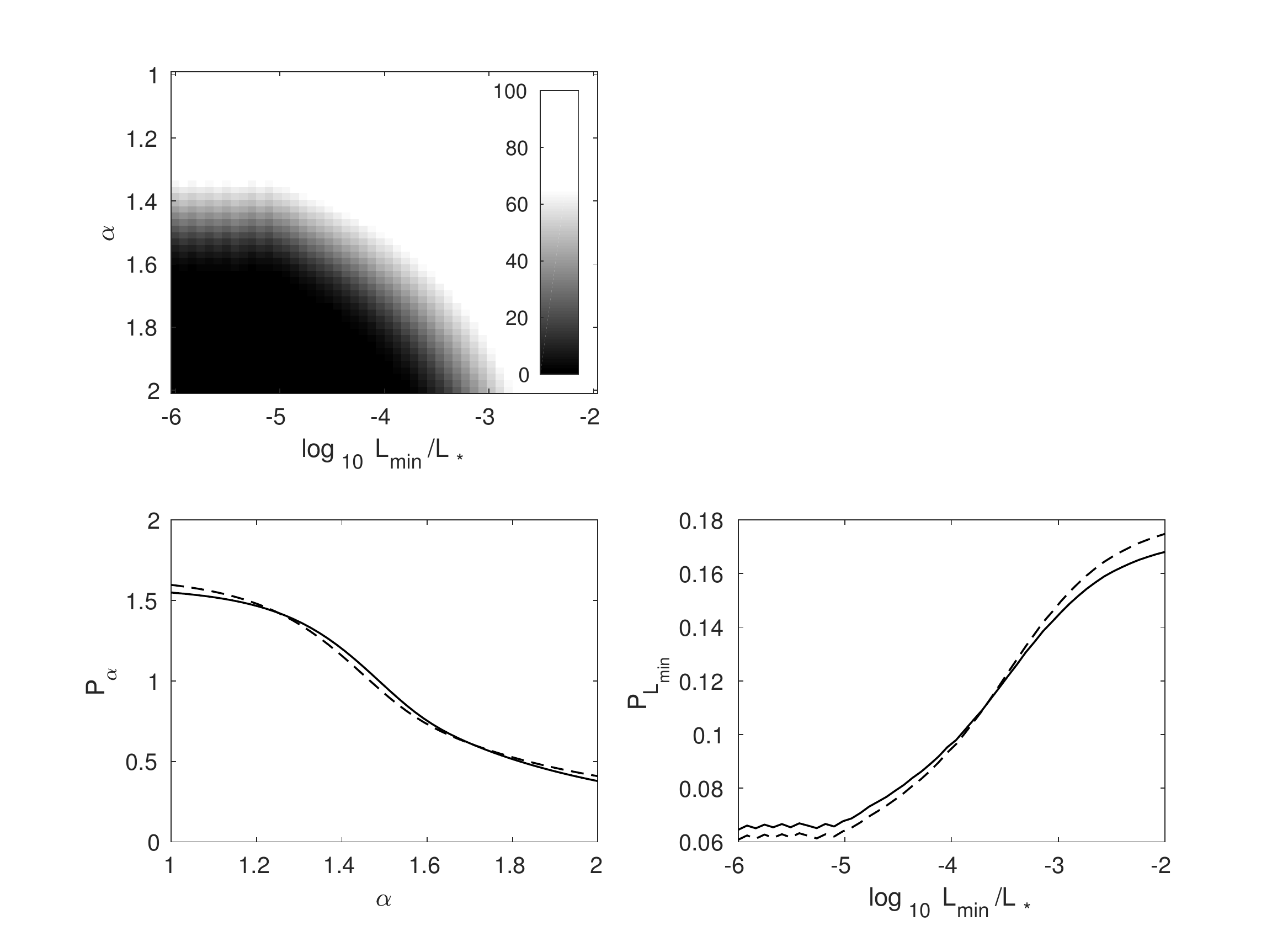}
\caption{Top: $P_0(\alpha, L_{\rm min}|\dot N_{\rm FRB})$, probability of detecting no FRBs from Virgo in 300 hours of ASKAP observations as a function of  $\alpha$ and $L_{\rm min}$ and assuming the total of $10^4$ FRBs per sky per day above a detection threshold of 1 Jy and at $z \leq 1$. Colour-code is shown on the colour bar with high probability of non-detection in white and low probability in black. Bottom: 1D-PDFs for $\alpha$ (marginalised over $L_{\rm min}$, left) and $L_{\rm min}$ (marginalised over $\alpha$, right). The 1D PDFs are shown in the case of the  SFR-driven FRBs (solid) and $M_*$ (dashed).} 
\label{Fig:Prob}
\end{center}
\end{figure*}
 
Next, we identify Virgo galaxies within the observed field (as specified in Fig.~\ref{fig:virgo_rosat}) using the online Virgo catalogue \citep{Kim2014}. Following \citet{Fialkov2018}, for each Virgo galaxy we calculate stellar mass using standard mass-luminosity relations \citep{Bernardi:2010} with luminosities extracted from the catalogue, and the SFR is calculated using the  SFR$-M_*$ relation \citep[e.g.,][]{Brinchmann:2004}. 
Including all the galaxies located within the field of view, we estimate the total expected number of FRBs from Virgo,  $\langle N_{\rm FRB}^{\rm Virgo}\rangle_{\alpha,L_{\rm min}|\dot N_{\rm FRB}}$, for a fixed value of $\dot N_{\rm FRB}$ and as a function of $L_{\rm min}$ and $\alpha$ using the pre-calculated normalisation coefficients,  $R^{\rm int}_{\rm SFR}$ and  $R^{\rm int}_{\rm M*}$. The predictions are shown in Fig.~\ref{Fig:NFRBs} (right panel) as a function of $L_{\rm min}$ and for three choices of $\alpha$ (1.3, 1.5, 1.7) and for $\dot N_{\rm FRB}=10^4$ FRBs per sky per day with flux $>1$ Jy and at $z\leq 1$. As anticipated, the lower is  $L_{\rm min}$, the more abundant is the population of faint detectable FRBs and the higher is  $\langle N_{\rm FRB}^{\rm Virgo}\rangle_{\alpha,L_{\rm min}|\dot N_{\rm FRB}}$. The number counts flatten at $L_{\rm min} = 6.4\times 10^{-6} L_*$ which corresponds to the sensitivity limit of ASKAP.  
As discussed above, it is likely that the detected FRB is behind Virgo as none of the galaxies from the Virgo cluster is located close to the line of sight. We estimate the probability to detect zero FRBs from Virgo, $P_0(\alpha, L_{\rm min}|\dot N_{\rm FRB})$, as a function of the model parameters using Poisson statistics with the expectation value of  $\langle N_{\rm FRB}^{\rm Virgo}\rangle_{\alpha,L_{\rm min}|\dot N_{\rm FRB}}$. 
Because of the high number counts of faint FRBs in the cases with steep luminosity functions and low values of $L_{\rm min}$, the probability for non-detection (the black region in the two-dimensional probability distribution, the top panel of Figure~\ref{Fig:Prob}) is low in these cases. Such scenarios are ruled out by the data presented in this paper.
On the other hand, in the cases with shallow luminosity function and high values of $L_{\rm min}$ the population is intrinsically bright. As a result, number counts from Virgo are low compared to the yield from the cosmological volume within the field of view. In such cases, it is more likely to find an FRB originating behind Virgo than within the cluster and $P_0(\alpha, L_{\rm min}|\dot N_{\rm FRB})$ is high (white region of the 2D PDF, the top panel of Figure \ref{Fig:Prob}).

Marginalising over one of the parameters we compute one-dimensional  PDFs for the other parameter (lower panels in Figure \ref{Fig:Prob}). Following the indication from \citet{Luo2018} we assume uniform prior on $\alpha$ within 1.2--1.8 and a uniform distribution in $\log_{10}L_{\rm min}$ over the range $[10^{-6}-10^{-2}]L_*$. We find that for the total of
$10^4$ FRBs per sky per day with flux $>1$ Jy and at $z\leq 1$, the non-detection of FRBs from Virgo is consistent with  $\alpha\leq 1.52 $ at 68\% confidence for both the SFR-driven case and the $M_*$-driven case.  We also find a lower limit on $L_{\rm min}$, with  $L_{\rm min}> 7.9\times10^{-5} L_*=1.6\times 10^{40}$~erg~s$^{-1}$  at 68\% confidence  for the SFR-driven case and with $L_{\rm min}> 9.5\times10^{-5} L_* = 1.9\times 10^{40}$~erg~s$^{-1}$  for the $M_*$-driven case.  For 10$^3$ FRBs per sky per day with flux $> 1$ Jy and at $z\leq 1$, the constraints are weaker (because of the lower number counts expected). We find  $\alpha\leq 1.58$ and  $L_{\rm min}> 4.1\times10^{-5} L_* = 6.5\times 10^{39}$~erg~s$^{-1}$ (both at 68\% C.L.).

\section{Conclusions}
We have presented the discovery and follow-up observations of FRB~180417 from a targeted search of the Virgo cluster. The search was motivated by the discussion by \citet{Fialkov2018}, of possible enhancement in FRB rates in the direction of rich galaxy clusters. The FRB was followed up for 27 hours with 4 more sensitive telescopes at L-Band. No repeat bursts were detected from the target location. We also followed up the FRB in the optical band using the PROMPT5 telescope, but no sources were discovered.

We argue that the FRB is likely behind the Virgo cluster as the Galactic and intracluster DM contribution was less than the DM of the FRB. Assuming FRB 180417 is beyond Virgo, we constrain for the first time intrinsically faint FRBs ruling out scenarios with a steep faint-end slope of the luminosity function and extremely low values of the minimum intrinsic FRB luminosity. For the total of  $\dot N_{\rm FRB} = 10^4$ FRBs per sky per day above a threshold of 1 Jy and out to redshift of 1, the minimum luminosity has to be higher than $2\times 10^{40}$~erg~s$^{-1}$ at 6\% confidence level (and higher than $6.5\times 10^{39}$ erg s$^{-1}$ for  $\dot N_{\rm FRB} = 10^3$ FRBs per sky per day). The luminosity function has to be rather shallow with the slope of 1.52 or lower for $10^4$ FRBs per sky per day (and of 1.58 or lower for 10$^3$ FRBs per sky per day).

Our unique limits on the faint-end population of FRBs are enabled solely by the combination of the target cluster search and the large field of view and sensitivity of ASKAP. Blind searches with less sensitive instruments such as the Canadian Hydrogen Intensity Mapping Experiment (CHIME) (The CHIME/FRB Collaboration et al. 2019b,c), even though reveal a significant number of new FRBs, are detecting only very bright events. In such searches, the faint-end population remains unconstrained. Further FRB surveys of galaxy clusters with high-sensitivity instruments will shed more light on the minimum intrinsic luminosity of FRBs.
\section*{Acknowledgements}
We thank the anonymous referee for the insightful comments that signifcantly improved the original manuscript. D.A. and D.R.L. acknowledge NSF award AAG-1616042. All authors acknowledge support from the NSF awards OIA-1458952 and PHY-1430284. D.R.L. also acknowledges support from the Research Corporation for Scientific Advancement. A.F. is supported by the Royal Society University Research Fellowship.
W.F., C.F. and R.M.S. acknowledge support through  Australian Research Council (ARC) grants FL150100148. R.M.S. acknowledges salary support through grant CE170100004. K.W.B., J.-P. M., and R.M.S. acknowledge support through ARC grant DP18010085.
G.P. acknowledge support provided by the Millennium Institute of Astrophysics (MAS) through grant IC120009 of the Programa Iniciativa Cientifica Milenio del Ministerio de Economia, Fomento y Turismo de Chile.
S.O. is supported by the Australian Research Council grant FL150100148.
B.W.S and K.R. acknowledges support from European Research Council Horizon 2020 grant (no. 694745) during which part of this work was done. DL acknowledges the support from NSFC No. 11725313, No. 11690024, and CAS Strategic Priority Research Program No. XDB23000000.
This work was supported in part by the Black Hole Initiative, which is funded by a JTF grant.

The filterbank cut outs for the three beams are available at \url{http://astro.phys.wvu.edu/files/askap\_frb\_180417.tgz}.






\bibliographystyle{mnras}
\bibliography{virgo}



\bsp	
\label{lastpage}
\end{document}